# Deep Learning Estimation of Multi-Tissue Constrained Spherical Deconvolution with Limited Single Shell DW-MRI

Vishwesh Nath*[a], Sudhir K. Pathak*[b], Kurt G. Schilling[c], Walt Schneider[b,d], Bennett A. Landman[a,c]
[a]Electrical Engineering & Computer Science, Vanderbilt University, Nashville, TN
[b]Learning Research and Development Center, University of Pittsburgh, Pittsburgh, PA
[c]Radiology, Vanderbilt University Medical Center, Nashville, TN
[d]Department of Psychology, University of Pittsburgh, Pittsburgh, PA

## ABSTRACT

Diffusion-weighted magnetic resonance imaging (DW-MRI) is the only non-invasive approach for estimation of intra-voxel tissue microarchitecture and reconstruction of in vivo neural pathways for the human brain. With improvement in accelerated MRI acquisition technologies, DW-MRI protocols that make use of multiple levels of diffusion sensitization have gained popularity. A well-known advanced method for reconstruction of white matter microstructure that uses multi-shell data is multi-tissue constrained spherical deconvolution (MT-CSD). MT-CSD substantially improves the resolution of intra-voxel structure over the traditional single shell version, constrained spherical deconvolution (CSD). Herein, we explore the possibility of using deep learning on single shell data (using the b=1000 s/mm$^2$ from the Human Connectome Project (HCP)) to estimate the information content captured by 8$^{th}$ order MT-CSD using the full three shell data (b=1000, 2000, and 3000 s/mm$^2$ from HCP). Briefly, we examine two network architectures: 1.) Sequential network of fully connected dense layers with a residual block in the middle (ResDNN), 2.) Patch based convolutional neural network with a residual block (ResCNN). For both networks an additional output block for estimation of voxel fraction was used with a modified loss function. Each approach was compared against the baseline of using MT-CSD on all data on 15 subjects from the HCP divided into 5 training, 2 validation, and 8 testing subjects with a total of 6.7 million voxels. The fiber orientation distribution function (fODF) can be recovered with high correlation (0.77 vs 0.74 and 0.65) and low root mean squared error ResCNN:0.0124, ResDNN:0.0168 and sCSD:0.0323 as compared to the ground truth of MT-CST, which was derived from the multi-shell DW-MRI acquisitions. The mean squared error between the MT-CSD estimates for white matter tissue fraction and for the predictions are ResCNN:0.0249 vs ResDNN:0.0264. We illustrate the applicability of high definition fiber tractography on a single testing subject with arcuate and corpus callosum Tractography. In summary, the proposed approach provides a promising framework to estimate MT-CSD with limited single shell data. Source code and models have been made publicly available.

**Keywords:** Deep learning, Spherical Harmonics, DW-MRI, MT-CSD, Tractography

## 1. INTRODUCTION

Diffusion-weighted magnetic resonance imaging (DW-MRI), a non-invasive in-vivo MR imaging technique captures unique information regarding the microstructure of the human brain [1]. One of the first microstructure analysis techniques was diffusion tensor imaging (DTI) [2]. However, DTI has been limited by only recovery of fiber populations in a primary direction. Multiple advanced acquisition schemes with advanced reconstruction methods were proposed to detect crossing fiber populations [3]. The advanced reconstruction techniques are collectively referred to as high angular resolution diffusion imaging (HARDI) methods [4]. A primary application of reconstructed microstructure is for constructing the white matter (WM) neural pathways of the human brain also known as tractography [5]. Advanced tractography methods, such as high definition fiber tractography (HDFT) [6], have been applied for utilized for neurosurgery guidance. A caveat is that HDFT requires multi-shell DW-MRI (multiple diffusivity values) acquisitions which are expensive and take much more time as compared to a single shell acquisition [6]. This work is focused towards recovery of HDFT with single shell DW-MRI acquisitions (Fig 1).

There are multiple methods that can be used as a prior for HDFT such as generalized q-sampling (GQI) [7] and multi-tissue constrained spherical deconvolution (MT-CSD) [8] both of which are microstructure reconstruction methods. This work tackles the problem of microstructure reconstruction only using the single shell DW-MRI acquisitions. There are a couple of existing approaches that have shown the possibility of recovery of tissue volume fraction from single-shell data. The first poses it as a non-negative factorization problem [9]. The second is a deep learning approach which directly takes the input of diffusion weighted images for fiber orientation distribution function (fODF) reconstruction [10]. The first approach has only been shown to reconstruct microstructure while the second one is restricted in terms of broader applicability as it is tied to input of diffusion weighted images directly and does not include a joint estimation of tissue fraction. Herein, we propose to use deep learning techniques which ensure broader applicability by the usage of spherical harmonics and we explore the differences between using single voxel and cubic patches for reconstruction of MT-CSD.

Deep learning has become a powerful tool for learning non-linear mappings between a set of inputs and outputs where a non-linear mapping exists [11]. Although deep learning has been quite useful in other medical imaging domains, it is still in its nascent stages for DW-MRI. Recent work has been seen in microstructure estimation, harmonization and k-space reconstruction [12-14]. For this specific problem, we explore two different network architectures for recovery of MT-CSD microstructure. The first approach is a residual deep neural network (ResDNN) [15] of five layers with a residual block in between and takes input of a single voxel in the form of spherical harmonics (SH) while providing the output of fODF derived from MT-CSD which are also in the form of SH. The second network takes an input of a cubic patch of voxels and makes the prediction of the center voxel thus using spatial information as features for the deep learning network. This network has five convolutional layers with a residual block further consisting of convolutional layers (ResCNN). This network is inspired from previous work [16].

The methods proposed have been trained, validated, and tested on the human connectome dataset [17]. Deep learning networks were trained on 5 subjects and validated on 2. While 8 subjects were withheld for testing. The deep learning methods were also compared with the silver standard of super-resolved constrained spherical deconvolution (sCSD) [18] as a baseline. All comparisons were made with the MT-CSD being considered as the ground truth.

## 2. DATA & METHODS

### 2.1 Human Connectome Project Data

The human connectome project (HCP) dataset has an advanced acquisition scheme with three different diffusivity values 1000, 2000 and 3000 s/mm$^2$. All three diffusivity values are acquired with 90 gradient directions with interspersed b0's. The pre-processed dataset provided was used for this work. All diffusion weighted volumes were normalized by the mean b0 as a standard pre-processing step. A total of 15 HCP subjects with the above acquisition scheme were used (Training: 5, Validation: 2, Testing: 8).

For training the proposed networks only the single shell of diffusivity value at 1000 s/mm$^2$ was used. The diffusion weighted volumes of that specific shell were fitted to 8$^{th}$ order SH which will be utilized as input training data for the network. Spherical harmonics in context of DW-MRI have become a standard way for representation of data with minimal representation error [19]. The output for the network which can be broken into two different parts 1.) SH coefficients of the fODF (8$^{th}$ order) which are reconstructed using MT-CSD on all three shells (1000, 2000 & 3000 s/mm$^2$) of DW-MRI data. 2.) Tissue volumes fractions which are scalar values for cerebrospinal fluid, apparent fiber density (white matter fraction), gray matter fraction.

### 2.2 Deep Learning Networks

The ResDNN is inspired from prior work and consists of five full connected dense layers (Fig 2). The number of neurons used per layer are x1: 400, x2: 45, x3: 200, x4: 45, x5: 200. The residual block is formed using the addition of the x2 and x4 layer. All layers are activated using 'relu'. The inputs are a vector of 1x45 coefficients of 8$^{th}$ order DW-MRI SH. The outputs are 1x45 fODF SH coefficients at 8$^{th}$ order with an additional 1x3 scalar vector which represents the tissue fraction volume. To adapt the use of fractional volumes as output we use a modified loss function which is defined in (1) where m denotes the number of samples $y_{true}$ is the set of fODF SH derived from MT-CSD $y_{pred}$ is the set of SH predictions made by ResDNN $P_{true}$ denotes the vector of tissue fraction value while $P_{pred}$ denotes the predicted vector of tissue fractions.

$$L = \frac{1}{m}\sum_{i=1}^{m} \alpha(y_{true_i} - y_{pred_i})^2 + \beta (P_{true_i} - P_{pred_i})^2$$

The ResCNN architecture stems from prior work where the network was originally intended for a harmonization problem. The network takes an input of 3x3x3x45 where the cubic patch consists of $8^{th}$ order DWMRI SH coefficients for each voxel in the cubic patch (Fig 2). The output is the same as defined for the ResDNN. This network architecture is divided into three parts, its core part being the residual block which consists of multiple functional units (each functional unit is dedicated to a specific order of SH). The residual block can be stacked multiple times keeping the spatial dimensions intact. For our purpose we use a single block. The residual block is connected with two more convolutional kernels which is finally connected to a dense layer for predicting the center voxel of the cubic patch. All layers are 'relu' activated. We use the same modified loss as described above for ResDNN.

**2.3 Evaluation Criteria**

To compare the predictions of the proposed deep learning methods we use angular correlation coefficient (ACC) [20] to evaluate the similarity of the prediction when compared with the ground truth estimate of MT-CSD. ACC is a generalized measure for all fiber population scenarios. It assesses the correlation of function of all directions over a spherical harmonic expansion. In brief, it provides the estimate of how closely a pair of fODF's are related on a scale of -1 to 1 where 1 is the best measure. Here 'u' and 'v' represent sets of SH coefficients.

$$ACC = \frac{\sum_{j=1}^{\infty}\sum_{m=-j}^{j} u_{jm} v_{jm}^*}{\left[\sum_{j=1}^{\infty}\sum_{m=-j}^{j}|u_{jm}|^2\right]^{0.5} \cdot \left[\sum_{j=1}^{\infty}\sum_{m=-j}^{j}|v_{jm}|^2\right]^{0.5}} \quad \ldots (2)$$

The tissue fraction volumes are assessed using mean squared error (MSE) per voxel and per tissue fraction. MSE has been evaluated over the entire brain volume in encompassing all regions of the brain.

## 3. RESULTS

We observe that ResCNN shows the most skewed distribution towards high correlation (Fig 3A) followed by ResDNN and sCSD. The mean ACC values across all 8 subjects were found to be 0.67, 0.72 and 0.64 for ResCNN, ResDNN and sCSD. Following subject wise ACC distributions (Fig 3B, 3C, 3D) for the entire brain volume we can see that ResCNN shows the most skewed distribution towards higher ACC for all subjects, followed by ResDNN and sCSD. Non-parametric signed rank test for all pairs of distributions were found to be p < 0.001. The root mean squared error across all pairs voxels for all subjects was found to be sCSD: 0.0323, ResDNN: 0.0168, ResCNN: 0.0124. Visually, the spatial map (Fig 4) of the middle axial slice for a single subject indicates that sCSD has high ACC for WM, while ResDNN shows improvement in the circulatory regions between WM and GM with ResCNN showing the highest ACC. All three methods show low ACC for CSF and GM regions.

The RMSE for CSF and GM is lower for ResDNN as compared to ResCNN (Table 1). While ResCNN only exhibits lower RMSE for only WM. The spatial maps for the middle axial slice of the same subject indicate higher error (MSE) of CSF and GM for ResCNN as compared to ResDNN. However lower MSE can be observed for WM for ResCNN as compared to ResDNN. Non-parametric signed rank test between distribution of MSE errors for the pair per tissue fraction were found to be p < 0.001.

## 4. DISCUSSION

In this work, we depict a full pipeline that begins from scanner acquisition to HDFT reconstruction using only a single shell DW-MRI acquisition. Deep learning on SH coefficients allows for generalizability and applicability when other scanner acquisitions need to be used for HDFT reconstruction. The deep learning approaches proposed for this work well and could be further improved by exploration of more intricate deep learning architectures and larger training datasets. While we have shown the proposed method to be applicable at a clinical diffusivity value, further validation on reduced numbers of gradient directions is necessary.

To do a full 3 shell by 90 direction scan takes ~40 minutes. The deep learning can potentially provide usable data for HDFT quality scans at ~5 minutes for a diffusivity value of 1000 s/mm$^2$ with 32 gradient directions. Clinical MRI use of advanced methods is very limited due to longer scan acquisition time. Reducing scan time by a factor of 8 could set the conditions for clinical use of HDFT quality scanning to grow dramatically perhaps several orders of magnitude.

Table 1. RMSE of all tissue fraction volumes.

| Method | RMSE CSF | RMSE GM | RMSE WM |
|---|---|---|---|
| ResDNN | **0.0135** | **0.0305** | 0.0264 |
| ResCNN | 0.0173 | 0.0320 | **0.0249** |

# ACKNOWLEDGEMENTS

This work was supported by R01EB017230 (Landman). This work was conducted in part using the resources of the Advanced Computing Center for Research and Education at Vanderbilt University, Nashville, TN. This project was supported in part by the National Center for Research Resources, Grant UL1 RR024975-01, and is now at the National Center for Advancing Translational Sciences, Grant 2 UL1 TR000445-06. This research was supported in part by the Intramural Research Program, National Institute on Aging, National Institutes of Health. The content is solely the responsibility of the authors and does not necessarily represent the official views of the NIH. This work has been supported by Nvidia with supplement of hardware resources (GPU's) in the form of a Titan Xp. This work was supported by the U.S. Army Medical Research and Material Command and from the U.S. Department of Veterans Affairs Chronic Effects of Neurotrauma Consortium under Award No. W81XWH-13-2-0095. The U.S. Army Medical Research Acquisition Activity, and the Chronic Effects of Neurotrauma Consortium/Veterans Affairs Rehabilitation Research & Development project F1880, US Army 12342013 (W81XWH-12-2-0139), Office of Naval Research and Naval Health Research Center (W911QY-15-C-0043).

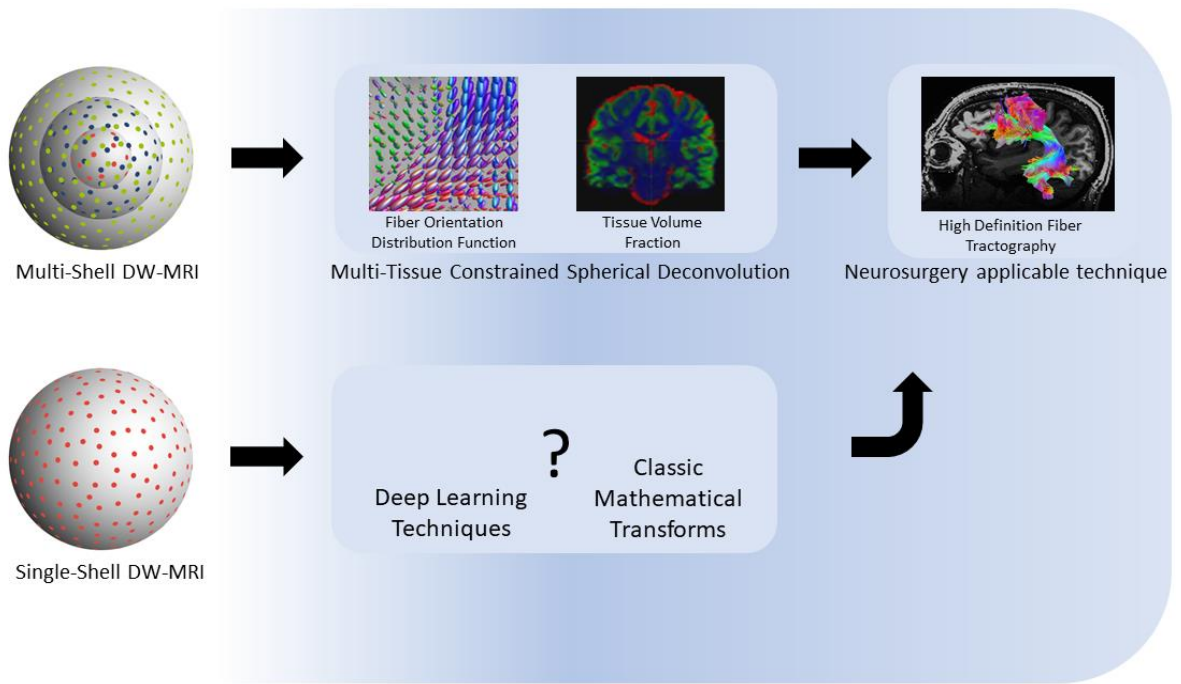

Figure 1: The top row describes the pipeline of obtaining microstructure information from multi-shell DW-MRI which in turn can be used to perform high definition fiber tractography. The bottom row describes the problem that we tackle for this work where we explore of how to perform high definition fiber tractography using only single-shell DW-MRI data.

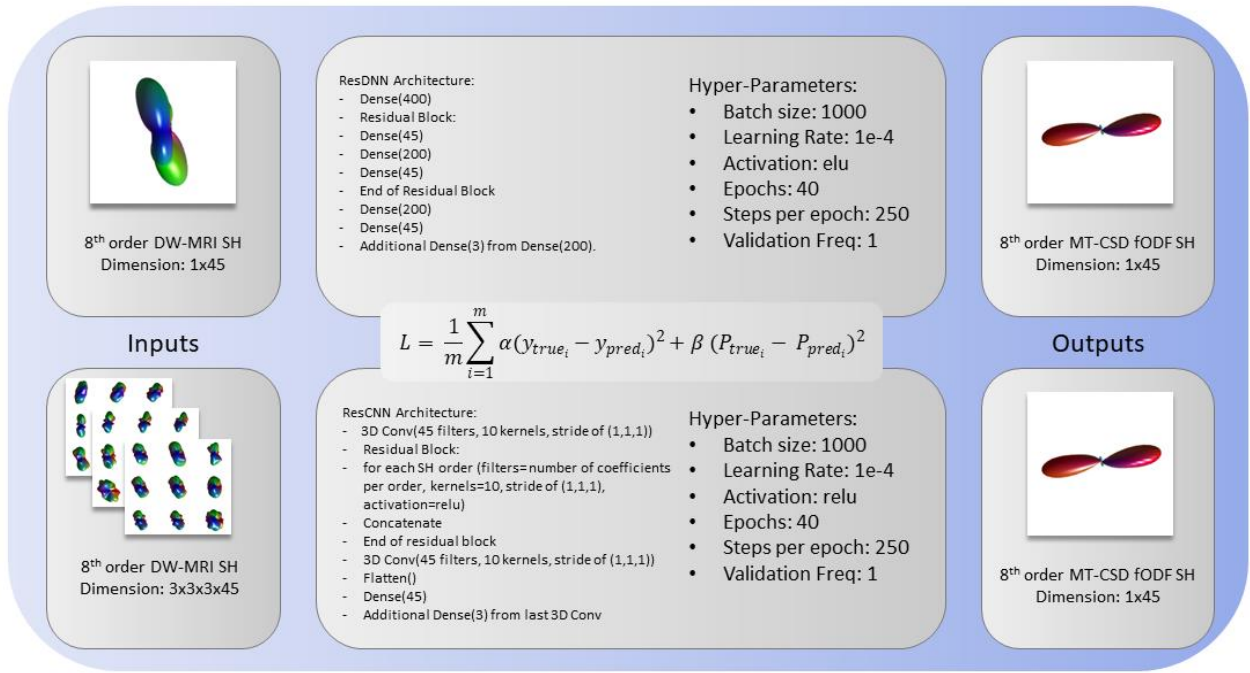

Figure 2: Top row describes the input, the architecture and hyper-parameters of ResDNN and then the output of the network with the loss function. The bottom row descirbes the same for ResCNN architecture.

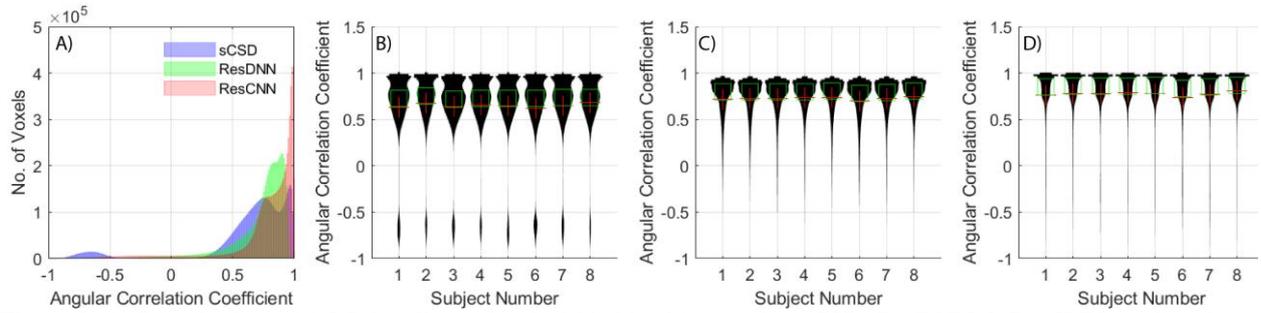

Figure 3: A) Distribution of ACC for all voxels of withheld subjects using sCSD, ResDNN & ResCNN predicted fODF when compared with fODF of MT-CSD. B) Subject wise distribution of ACC for sCSD. C) Subject wise distribution of ACC for ResDNN. D) Subject wise distribution of ACC for ResCNN.

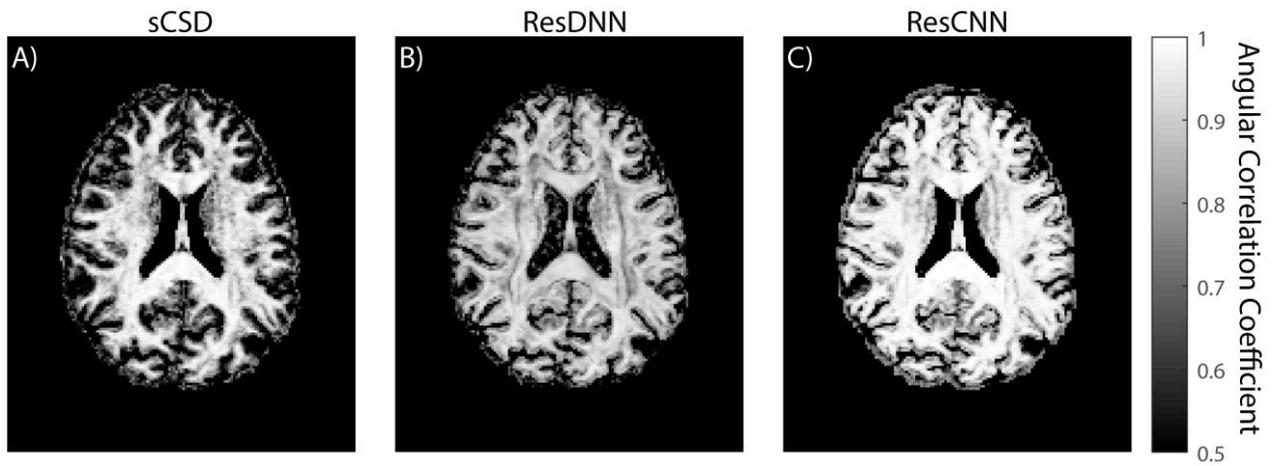

Figure 4: Spatial angular correlation coefficient maps of the middle axial slice of the brain of a single subject. A) fODF predictions from sCSD. B) fODF predictions from ResDNN. C) fODF predictions from ResCNN

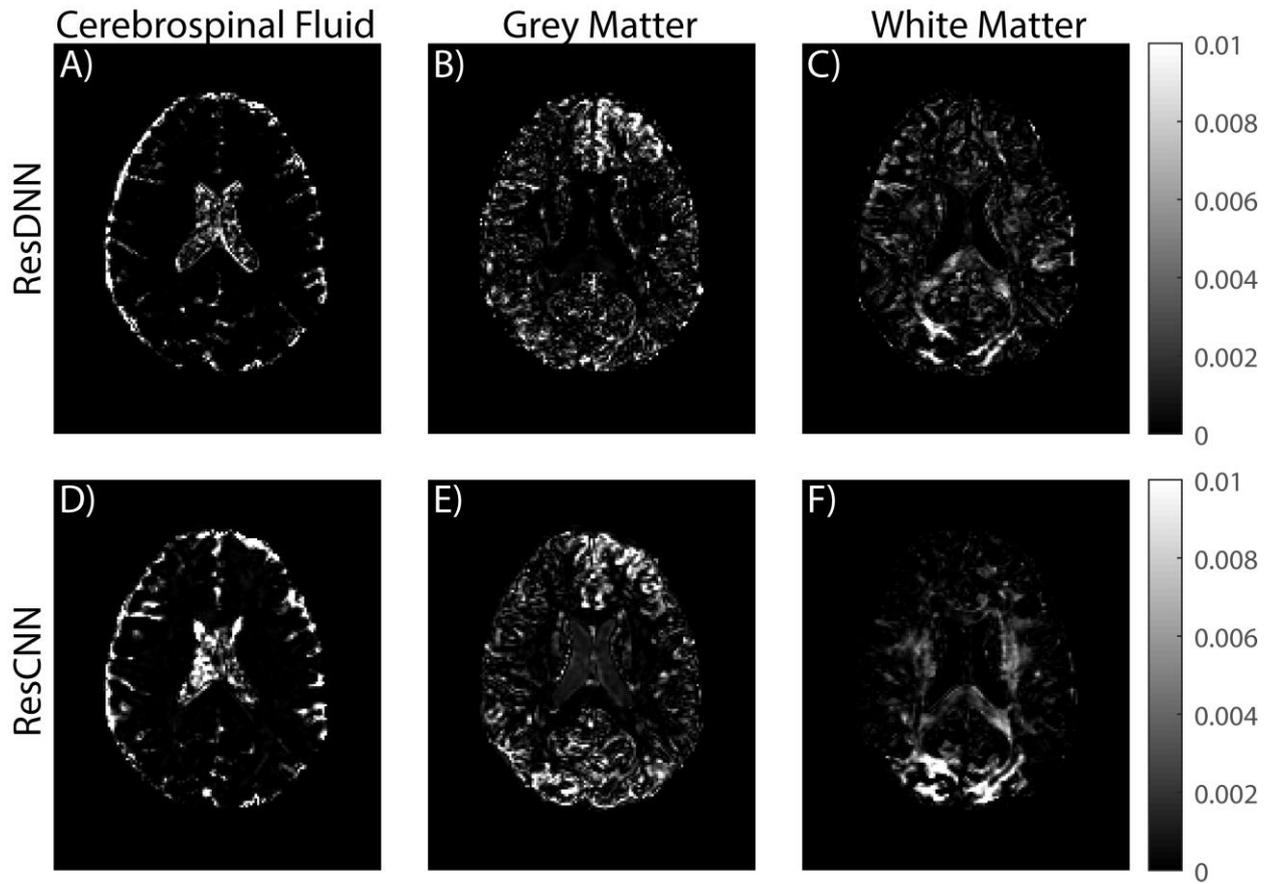

Figure 5: Spatial maps of mean squared error between tissue fraction estimates of ResDNN and ResCNN with estimates of MT-CSD.

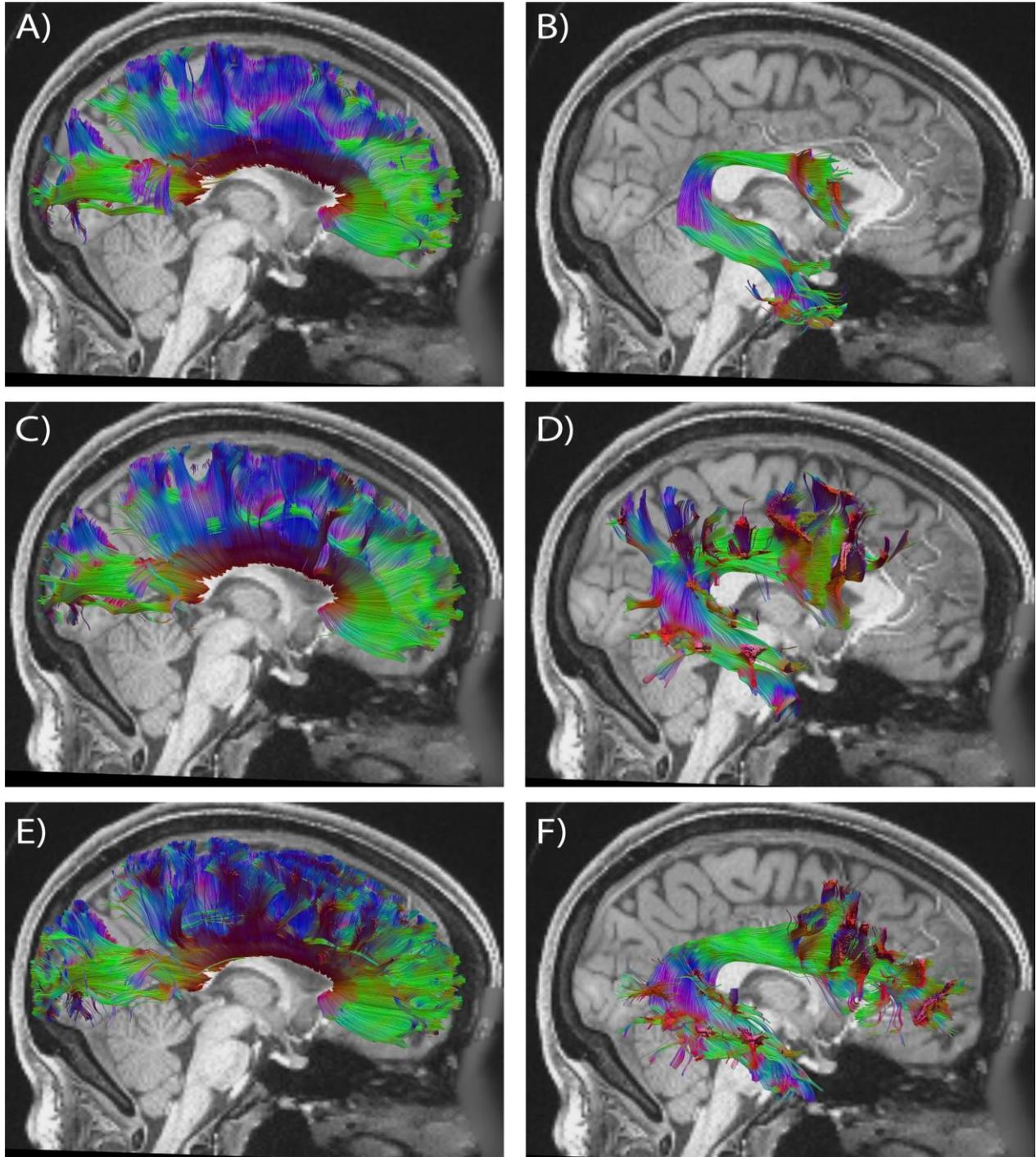

Figure 7: Tractography reconstruction of HDFT for the tract corpus callosum using A) ResDNN C) ResCNN E) MT-CSD. Reconstruction of HDFT for tract arcuate using B) ResDNN D) ResCNN and F) MT-CSD.

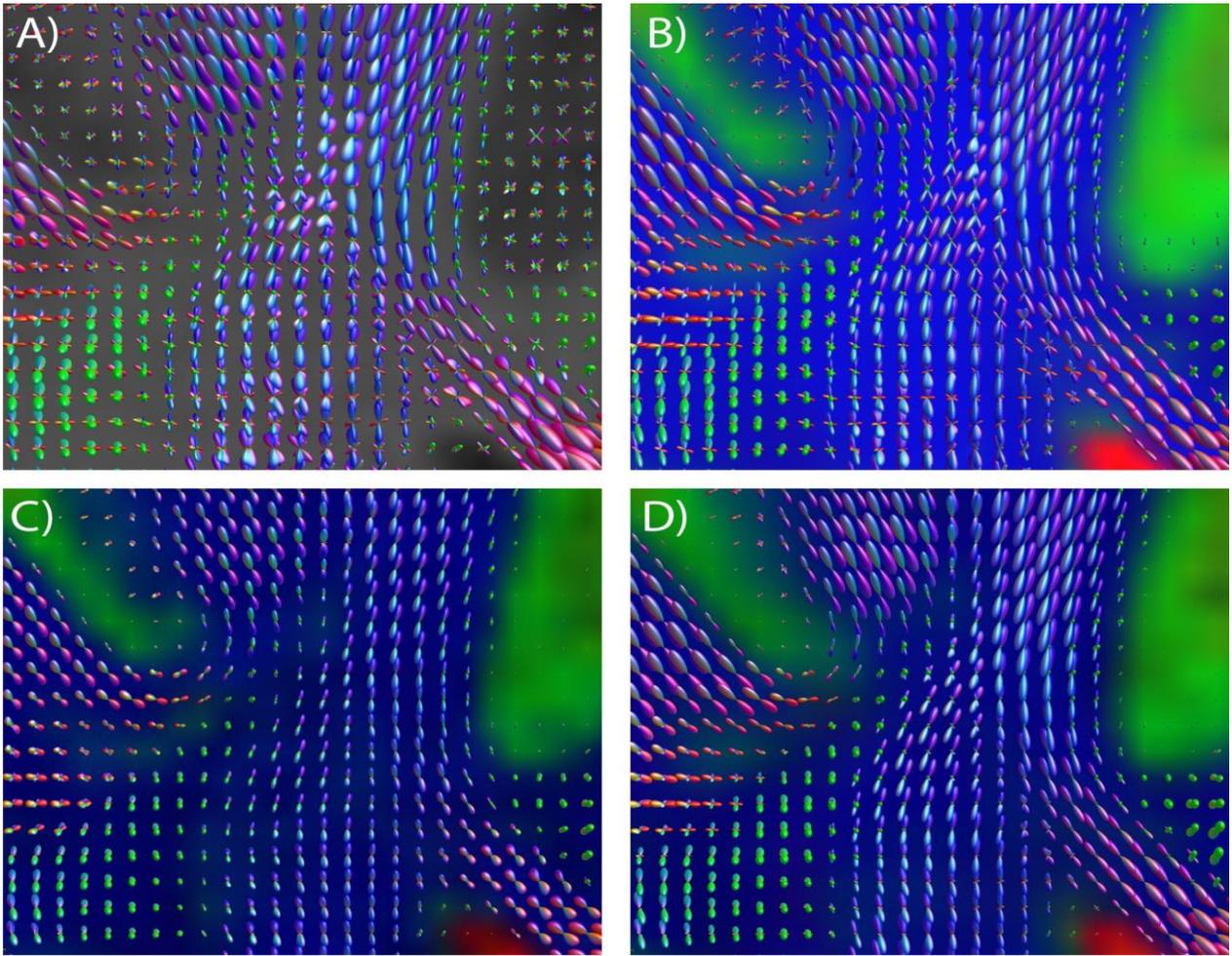

Figure 6: Crossing fiber region of interest of fODF on a middle coronal slice reconstruction shown for A) sCSD, C) ResDNN, D) ResCNN and B) MT-CSD.